\documentclass[aps,floatfix,superscriptaddress,notitlepage,nofootinbib]{revtex4-1}
\usepackage[utf8]{inputenc}
\usepackage{svg}
\usepackage{amsmath,amssymb,amsfonts,graphics,graphicx,dcolumn,bm,enumerate}
\usepackage{comment,natbib,appendix}
\usepackage{multirow,color}
\usepackage{chngpage}
\usepackage{afterpage}
\usepackage{xcolor}
\usepackage{amsthm}
\usepackage{natbib}
\usepackage{hyperref}
\usepackage[margin=1.2in,headheight=40pt,headsep=0.7in]{geometry}
\usepackage{epstopdf}
\usepackage{float}
\usepackage{natbib}
\usepackage{soul}
\usepackage{setspace}

\newcommand{\isi}
{\affiliation{Economic Research Unit, Indian Statistical Institute, Kolkata 700108, India.}}

\begin{document}

\title{International Centre for the Advancement of Multidisciplinary Studies on Socio-Economic Systems}

\author{Suchismita Banerjee}
\email[Email: ]{suchib.1993@gmail.com}
\isi

\author{Manipushpak Mitra}
\email[email: ]{mmitra@isical@ac.in}
\isi 

\begin{abstract}

\noindent We start by summarising very briefly the various prior attempts (during the last one and half a decade), some of which were made as independent research centres and others as visiting centres with extensive visiting programs for luminaries from various basic sciences (Mathematics, Physics, Biology, Economics, and Sociology) and students from various institutions around the world for such interdisciplinary fusion of ideas and researches. Additionally, we briefly discuss the efforts that our institute has made (without any visible success so far, as in the other attempts elsewhere).
We then emphasise the critical need for such an international centre to attract stalwarts in the basic disciplinary fields as well as interested students from around the world in order to comprehend the world's global socio-economic dynamics.

\end{abstract}

\maketitle

The study of inter-disciplinary subjects is found to be very helpful in order to understand various social and natural phenomena.
For subjects like astrophysics, biophysics, or geophysics, the amalgamation of two different streams was quite natural (being constituent parts of the same science faculty of the Universities) and has been very successful. 
For biophysics, the success has often been formally recognized publicly: For example, theoretical physicist Max Delbr\"{u}ck was awarded Nobel prize in Physiology and Medicine in 1969 \citep{delbruck_url} and the experimental physicist  Peter Mansfield was awarded the Physiology  and Medicine Nobel prize in 2003 \citep{mansfield_url}. 
Since Nobel Prizes are not offered for main subjects like Astronomy or Geology, one can not perhaps argue in a similar way for the success of astrophysics or biophysics. 
However, a look at the main research tools and topics in astronomy today \citep{astrophysics_url}, like the optical astronomy, infrared astronomy, ultraviolet astronomy, radio astronomy,  X-ray astronomy, gamma-ray astronomy, etc for planetary astronomy, Solar astronomy, stellar astronomy, galactic astronomy and extra-galactic astronomy or cosmological astronomy are all physics-based.
Similarly, the list \citep{geophysics_url} of the main research topics in geoscience or earthscience, like the phenomena of plate tectonics, evolution of mountain ranges, volcanoes, earthquakes, geomagnetism etc, atmospheric science of the troposphere, stratosphere, mesosphere, thermosphere, exosphere, etc, are all entirely physics-based. 
These clearly confirm the importance and success of the interdisciplinary fields like astrophysics and geophysics.

For econophysics, however, the scenario is quite different. 
Even the fundamental study of economics is frequently included in the humanities and arts departments of the various universities.
For example, Master of Arts or MA  degrees in economics are still offered by many distinguished institutions like the University of Kyoto \citep{1}, Manchester University \citep{2}, Boston University \citep{3} (or MPhil degrees from Universities of Cambridge \citep{4}, Oxford \citep{5}, etc). 
Economics is a subject to understand the interactive behavior of economical agents. 
Such kinds of interactions are not directly governed by any physical law but are typically found to follow the self-evolved game-theoretical interactions which are of multi-scale nature (the interaction between agents is different from the interaction between states, which is composed of many agents). 
As economics is inherently a ``many-body" or ``many-agent" phenomena (individual constituents can not evolve a market), the statistical physics of many-body interacting systems may provide a natural framework to study such a multi-scaled complex interaction phenomena in the purview of economics.
Despite this well established demonstrated power of statistical physics in understanding the concepts of collectively interacting systems, many economists have an aversion to mixing economics with physics.
Such a consideration is also, however, not applicable for some institutes like the Indian Statistical Institute, the London School of Economics, the New School of Economics, Huji Israel, etc., where these subjects are treated as parts of generalised science and the logical analysis of them is frequently practised.
But such situations have so far been very rare.
Moreover, in most institutes and universities, there is a natural segregation observed among the students and researchers of these two fields, and the scope of interaction among inter-field students is very small.
Owing to such reasons the funding for the establishment of econophysics as a self-consistent interdisciplinary subject is observed to be very minimal.
Additionally, the available funding for science and technology could not have been used for such interdisciplinary research.
As a result, the growth of these interdisciplinary subjects has been slower than that of other interdisciplinary subjects.
Recently, however,  theoretical physicist like Doyne Farmer and economist Duncan Foley advocated strongly \citep{farmer_2009} in favor of  Agent-Based kinetic-exchange modelling for the study of economic systems and their evolving dynamical manifestations at global-level consequences and Doyne Farmer initiated an Agent-Based Modelling research program \citep{farmer} ``ABM research program in the Institute for New Economic Thinking at the Oxford Martin School".

\begin{figure}
    \centering
    \includegraphics[scale=0.6]{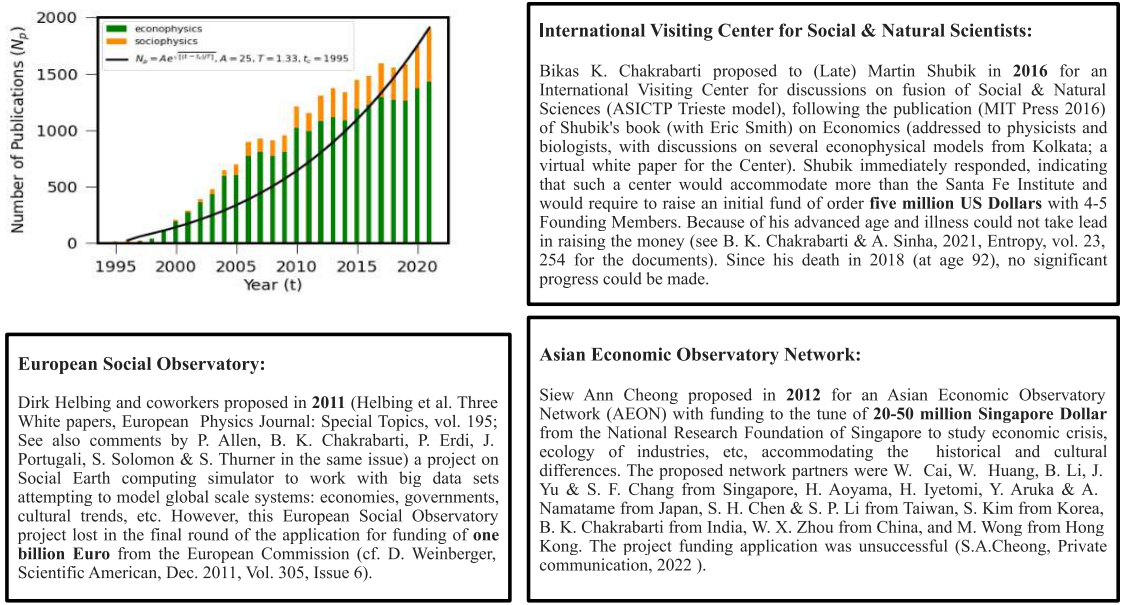}
\caption{Upper left pannel: the bar-plot of the number of publications ($N_{p}$) of the subjects, Econophysics (green) and Sociophysics (orange), throughout the years (1995 $\leq t \leq$ 2021). Black line indicates a sub-exponential fit to the data. Bottom left box: A failed attempt of the proposed project in 2011 (see \citep{Helbing}, \citep{Comments} \& \citep{Weinberger}). Right bottom box: Another unsuccessful project in 2012. Upper right box: One other attempt in 2016 which has also been unfruitful till date (see \citep{Chakrabarti}).}
\label{fig:1}
\end{figure}

Despite all the difficulties stated above, we can observe, from 1995 to 2021 (from Google Scholar data updated by November 2022), a substantial growth of Econophysics and Sociophysics (See Fig.~\ref{fig:1}).
The growth is typically observed to be following a sub-exponential morphology, but the situation can perhaps be drastically improved if we pay some attention to the funding issue that can potentially induce more interaction across social science and physics.
In addition, physicists are increasingly being acknowledged as possible significant contributors to the understanding of the complex dynamics of economies.
Though rare still, some  distinguished fellowships for the economists are now slowly being awarded to physicists in order to  recognize their contributions in economics (see e.g., \citep{banglar_anirban} for the award of The World Academy of Science Fellowship in economics \& Sociology in 2022 going to an Indian physicist, Jawaharlal Nehru University, Delhi).
For the past eleven years, Leiden University, home of one of the first (1969) Economics Nobel Laureates, Jan Tinbergen, has offered an Econophysics course \citep{X}.
There are at least two designated ``Professor" positions in Econophysics: Tiziana de Matteo, Professor of Econophysics, King's college, University of London \citep{Z} and Diego Garlaschelli, Associate Professor of Econophysics, University of Leiden \citep{A}.
Of course, many more physicists, mathematicians, and computer scientists are working actively in Econophysics and Sociophysics.
It may also be mentioned that the world’s oldest physics society \citep{S}, German Physics Society, instituted the annual `Young Scientist Award for Socio and Econophysics’ in 2002 and some really illustrious scientists \citep{D} in social and complexity sciences received this prize in the last two decades.

In view of all these, we would like to propose for a global initiative to create a centre for Econophysics and Socio-physics study where scientists from different disciplines across different countries might come and interact. 
In fact, the earlier (UNESCO) project, namely the Abdus Salam International Centre for Theoretical Physics in Trieste, may serve as a significant practical example in this regard.
Such an endeavour might include a robust visitor program, interdisciplinary schools and workshops, and the creation of ``source books" or even ``textbooks" (in addition to data banks, etc.) for these developing interdisciplinary knowledge domains.
In 2011, 2012 and 2016 different scientists throughout the world took some initiatives for some Global centres for interdisciplinary field (See Fig.~\ref{fig:1}) but all of them were unsuccessful projects.
The lack of a ``global" appreciation to develop such interdisciplinary centers for research, as well as the evolution of the subject and format of investigations in such an ``interdisciplinary" field, have been the primary reasons for the failure or slow pace of achieving the objectives of prior endeavours.
We believe, however, though the respective funding agencies for the two observatory proposals (mentioned in Fig.~\ref{fig:1} and discussed earlier) might have appreciated the academic and cultural aspects of these proposals, they rejected them perhaps because there was not enough evidence of immediate applicability to address the major social and economic problems at hand.
Another factor contributing to such failures in those days is probably the fact that the initiative involved only a few passionate scientists.

In another instance involving our own center (Indian Statistical Institute, ISI), the ``Policy Planning and Evaluation Committee" (PPEC)\citep{PPEC} of the ISI, in its June 22 (2011) meeting, evaluated a ``proposal for creating a Center for Econophysics and Quantitative Finance Research" and suggested that ``PPEC recognizes this to be an important proposal, but considering the availability of manpower and the current focus of ERU (Economic Research Unit), it recommends that the proposal be carried as a plan research project, but not as a full-fledged centre at this point of time.
However, the recruitment of faculty members in the area of Econophysics or related disciplines may be made in ERU if need.''
We are also pleased to know that comparable initiatives are being made in other significant institutions across the nation.

There are some success stories too like the Vienna-based Complexity Science Hub (CSH) \citep{CSH} that aims to advance ``Austria's research into complex systems, system analysis, and big data science."
The CSH was established in 2015 as a cooperative endeavour to promote big data science for societal good and to raise the profile of Austrian complexity research abroad.
The collaboration was further extended to several European centers in particular Vienna University of Economics and Business in Vienna, the Central European University in Budapest, the Santa Fe Institute in New Maxico.

Indeed, beginning in the third decade of the last century, Meghnad Saha had thought about the scientific, in particular statistical physical foundations of many social issues. 
His critical attention must have been drawn to the socioeconomic (income or wealth) inequalities.
Together with Biswambhar Nath Srivastava, he wrote the textbook ``Treatise on Heat" (first edition in 1931) \citep{F}, in which they urge the students to apply kinetic theory to obtain the Gamma-like income distribution depicted in Fig. 6 on page 105 discussing the Maxwell-Boltzmann energy distribution in ideal gases (see e.g., \citep{G}). 
Also, in the Princeton University, Princeton, the research in Social Physics was started in 1955 by John Q. Stewart in the physics department \citep{C}.
A reported summary of their social physics project says, ``In the early phase of their efforts, Professor Stewart and his colleagues in this enterprise confined their efforts to mass human relationships. 
They treated large aggregates of individuals as though they were composed of social molecules, without attempting to analyze the behavior of each molecule. 
They then attempted to describe demographic, economic, political, and sociological situations in terms of such physical factors as time, distance, mass (or molecular weight), and numbers of people (or number of atoms in the molecule)."
However none of these efforts were pursued for long.

However, in California Institute of Technology, California, the Division of Physics, Mathematics and Astronomy has added in their Core Curriculum Learning Outcomes \citep{B} ``Significant study in the humanities and social sciences. Students will be able to: (i) Explore and expand upon learning in fields beyond intended areas of specialization. (ii) Appreciate and understand the contributions of the humanities, social sciences, and arts to human endeavors. (iii) Engage in informed analysis of cultural, political, and economic issues."

Due to lack of funding and interactions among the experts on these subjects, the growth of subjects like econophysics and sociophysics might not have been at the intended level.
The growth can be increased by building a global visiting centre where researchers can temporarily stay and interact with the visiting members.
Here our intention is not to built a theoretical center but an observatory where people from all over the globe can visit and interact with each others.
Not only physics and economics experts, our target is to bring interested people from different subjects like management, data science.
The center will provide accommodation and hospitality to the visitors for the span of their visit only.
Additionally, there will only be a small number of permanent positions for the researchers.
The use of such a strategy will help to reduce the funds and at the same time provide a space for interaction between various researchers among the entire academic spectrum.
It is worth noting that, although there have been some failed attempts in the past, some indirect successes are also visible.
We believe, along with a model of governance and a sample budget, such a Centre needs a clear business plan.
Specifically, we require a list of advantages that would result from the proposed Centre for the hosting institution and the hosting nation (see the comments from Bruce Boghosian in subsection 3(C) of ref. \citep{Y}).

We further hope that the above-mentioned fund reduction strategy will be helpful in materialising such an international visitor center.
If this materialises, the field will receive enormous momentum and the society at large will greatly benefit from the increased expansion of its knowledge base.

\bigskip

\section*{Acknowledgements:}
We are grateful to Siew Ann Cheong for sending vital information regarding the Asian Economic Observatory Network and to Bikas K. Chakrabarti for some useful comments.

\end{document}